\documentstyle[12pt]{article}

\begin{document}
{\sf \begin{center} \noindent {\Large\bf Dynamo experiments in torsioned toroidal devices}\\[2mm]

by \\[0.1cm]

{\sl  L.C. Garcia de Andrade}\\Departamento de F\'{\i}sica
Te\'orica -- IF -- Universidade do Estado do Rio de Janeiro-UERJ\\[-3mm]
Rua S\~ao Francisco Xavier, 524\\[-3mm]
Cep 20550-003, Maracan\~a, Rio de Janeiro, RJ, Brasil\\[-3mm]
Electronic mail address: garcia@dft.if.uerj.br\\[-3mm]
\vspace{1cm} {\bf Abstract}
\end{center}
\paragraph*{}
Recently Shukurov, Stepanov and Sokoloff [\textbf{Phys. Rev. E 69
(2008)}] have suggested that Moebius flows can support dynamo
action. In this report, it is shown that a steady perturbation of a
magnetic field in a general twisted Riemannian flux tube may support
dynamo action. Instead of the twist number used in the above
reference, the focus here is on the Frenet torsion of the magnetic
flux tube. A relation between the constant torsion of the screw
dynamo torus, its internal radius, and the ratio between toroidal
and poloidal flow. Solution of self-induction equation for the screw
dynamo torsioned flow, can be solved to yield a Frenet torsion as
high as
${\tau}_{0}\approx{7.5{\times}{\Omega}^{\theta}Hz^{-1}m^{-1}}$, for
an applied random magnetic field of $<B^{0}>\approx{45 G}$ for an
induced steady perturbation of $B_{1}\approx{0.03G}$, as in the Perm
Riemannian torus experiment. The Moebius strip plays the role of the
propeller divertor which imprints the angular velocity
${\Omega}^{\theta}$, around the torus axis, to the dynamo flow, here
this role is played by torsion. Actually this situation is already
familiar to plasma physicists, where in stellarators torsion
substitutes very well the role played by external magnetic fields in
tokamaks making magnetic fields twist along plasma toroidal devices.
Weak torsion of the torus channel is assumed. Solution of the
equation for the random flow, yields a magnetic field maximum growth
rate of the order of
${\gamma}_{max}\approx{6{\times}10^{-2}{\Omega}^{\theta}}$. Just to
give an idea of how weak this is a galactic dynamo, may give rise to
a growth rate of ${\gamma}_{G}\approx{10^{-5}}s^{-1}$. In the
previous expression, one notes that the torsion contributes to
dynamo action. Actually a small torsion may be used to achieve
galactic dynamos growth rate. For magnetic Reynolds number of
$R_{m}\approx{16}$ the torsion is shown to act as a breaking
mechanism of the rotating Perm torus dynamo.{\bf PACS
numbers:\hfill\parbox[t]{13.5cm}{02.40.Hw. 47.65.-d,52.30.Cv.}}}
\newline
\section{Introduction}
 The investigation of the kinematic screw (helical) dynamo action
 \cite{1} in the flow of a cylindrical periodic tube with conducting wall, has been
 addressed by Dobler, Frick and Stepanov \cite{2}, by making use of
 an eigenvalue analysis of the steady regime, and the
 three-dimensional solution of the time-dependent self-induction
 equation. In the Perm dynamo experiment \cite{3,4,5} in Russia, a
 torus device is used. The first type of dynamo flow on a Riemannian
 toroidal space, where a uniform stretched flow presented dynamo
 action has been proposed for the first time by Arnold, Zeldovich,
 Ruzmaikin and Sokoloff \cite{6}. More recently two examples of
 dynamo action in Riemannian space \cite{7,8} has been presented.
 The first example was a stretch-twist and fold fast dynamo action in
 conformal Riemannian manifolds. The second example is application
 of the anti-fast-dynamo theorem by Vishik \cite{9} to the plasma
 devices. Yet more recently, Shukurov, Stepanov and Sokoloff
 \cite{10}, has been proposed a Moebius strip flow, may support
 dynamo action, where the natural twist of the Moebius strip is used
 to substitute the divertor fan of the Perm torus dynamo experiment,
 in the creation of the screw of the dynamo flow. This technique is
 actually similar to the one used by plasma physicists in the
 stellarators devices, where the presence of the Frenet torsion
 plays the role of external magnetic fields in tokamaks, to induce the existence of helical highly conducting magnetic fields. Since Frenet
 torsion of the magnetic flux tube is part of the twist \cite{11} of the
 tubes, a natural extension of the Moebius strip dynamo flow, would
 be to consider the torsioned dynamo flow action. In this paper such
 an enterprise is undertaken. As in the Moebius dynamo flow experiment proposal, here one assumes that the magnetic Reynolds $R_{m}$ is small as $R_{m}\approx{16}$. This paper is organized as follows:
 In section 2 the perturbed equation is solved and the equation between torsion of the dynamo flow and the ratio between toroidal and poloidal components of the flow is deduced.
 In section 3, the random time dependent self-induced equation is solved and the dynamo growth rate is determined. Section 4 presents discussions and conclusions.
 \newpage
\section{Screw dynamo flows in Riemannian space}
\vspace{1cm} Let us start this section, by defining the perturbed
random magnetic flow field $<\textbf{B}^{0}>$ as
\begin{equation}
\textbf{B}= <\textbf{B}^{0}>+\textbf{B}_{1}\label{1}
\end{equation}
where $\textbf{B}_{1}$ is the magnetic field stationary
perturbation, while $<\textbf{B}^{0}>$ is the random applied field.
By substitution of this expression into the self-induction
equation
\begin{equation}
{\partial}_{t}\textbf{B}=
{\nabla}{\times}(\textbf{v}{\times}<\textbf{B}>)+{\lambda}{\Delta}\textbf{B}\label{2}
\end{equation}
yields
\begin{equation}
{\partial}_{t}<\textbf{B}^{0}>={\gamma}<\textbf{B}^{0}>=
{\nabla}{\times}(<\textbf{v}{\times}\textbf{B}_{1}>)+{\lambda}{\Delta}<\textbf{B}^{0}>\label{3}
\end{equation}
where ${\lambda}$ is the diffusive coefficient,
$<\textbf{B}^{0}>=e^{{\gamma}t}B^{0}\textbf{t}$ and
${\Delta}={\nabla}^{2}$ is the Laplacian operator, which gradient
operator ${\nabla}$ is given in Riemannian curvilinear coordinates
by
\begin{equation}
{\nabla}=\textbf{e}_{r}{\partial}_{r}+\textbf{e}_{\theta}\frac{1}{r}{\partial}_{\theta}+\textbf{t}{\partial}_{s}\label{4}
\end{equation}
where thin tube metric can be obtained from the Riemannian metric of
the twisted magnetic flux tube \cite{8}
\begin{equation}
dl^{2}= dr^{2}+r^{2}d{{\theta}_{R}}^{2}+K^{2}(r,s)ds^{2} \label{5}
\end{equation}
by taking $K(r,s)=(1-{\kappa}r \cos{\theta}):=1$ , where  ${\kappa}$
is the Frenet curvature and the twist transformation angle is given
by
\begin{equation}
{\theta}(s):={\theta}_{R}-\int{{\tau}(s)ds} \label{6}
\end{equation}
one obtains the Riemannian line element of the thin flux tube
\begin{equation}
dl^{2}= dr^{2}+r^{2}d{{\theta}_{R}}^{2}+ds^{2} \label{7}
\end{equation}
which gives rise the above gradient del operator ${\nabla}$. Along
with the Frenet frame $(\textbf{t},\textbf{n},\textbf{b})$ one is
able to solve the above equations. The perturbation first order
equation becomes \cite{12}
\begin{equation}
{\partial}_{t}<\textbf{B}_{1}>={\gamma}<\textbf{B}^{0}>=
{\nabla}{\times}(\textbf{v}{\times}\textbf{B}_{1}-<\textbf{v}{\times}\textbf{B}_{1}>)+{\lambda}{\Delta}\textbf{B}_{1}
 \label{8}
\end{equation}
The other random equation can be expressed in operator form as
\begin{equation}
({\gamma}-{\lambda}{\Delta})<\textbf{B}^{0}>=
{\nabla}{\times}(<\textbf{v}{\times}\textbf{B}_{1}>)\label{9}
\end{equation}
As one shall see in the next section $\textbf{B}_{1}$ can be
expressed in terms of $<\textbf{B}^{0}>$, which shows that the last
equation is an eigenvalue equation. Actually the dynamo operator
\begin{equation}
{L}={\gamma}-{\lambda}{\Delta}\label{10}
\end{equation}
has been studied in compact Riemannian manifolds by Chiconne and
Latushkin \cite{13}. In the next section one shall solve this
equation to obtain the values of the growth rate. The relation
between ${\lambda}$ and $R_{m}$ is given by
\begin{equation}
{\lambda}=\frac{vl}{R_{m}}\label{11}
\end{equation}
where v and l are respectively the typical velocity and length
scales involved in the dynamo twisted torus experiment. Note here
that, as in the mean-field-magnetohydrodynamics \cite{14}, the
random vector magnetic field here
$<\textbf{B}^{0}>=<B^{0}>\textbf{t}$. In this section one can write
the equation for the perturbed field by assuming that it is a steady
perturbation or ${\partial}_{t}\textbf{B}_{1}=0$ which reduces the
self-induction equation to
\begin{equation}
{\Delta}\textbf{B}_{1}=-\frac{1}{\lambda}(<\textbf{B}^{0}>.{\nabla})\textbf{v}
\label{12}
\end{equation}
By computing the Laplacian operator ${\Delta}$ in Riemannian space
yields
\begin{equation}
{\Delta}=[{{\partial}_{r}}^{2}+(1-\frac{{{\tau}_{0}}^{2}}{r^{2}}){{\partial}_{s}}^{2}-{{\tau}_{0}}\cos{\theta}{\partial}_{r}+
\frac{1}{r}(\sin{\theta}-\cos{\theta}){\partial}_{s}]\label{13}
\end{equation}
where ${\tau}_{0}$ is the constant Frenet torsion of screw dynamo.
Splitting the magnetic fields into its toroidal and poloidal
components yields
\begin{equation}
\textbf{B}_{1}={B}_{s}\textbf{t}+{B}_{\theta}\textbf{e}_{\theta}
\label{14}
\end{equation}
Assuming that
${\partial}_{s}B_{s}=B_{r}={\partial}\textbf{B}_{1}=0$, and applying
this Riemannian operator ${\Delta}$ into the equation (\ref{12}),
after a long computation yields
\begin{equation}
{\Delta}\textbf{B}_{1}=[(1-\frac{{{\tau}_{0}}^{2}}{r^{2}}){{\partial}_{s}}^{2}\textbf{B}_{1}+
\frac{1}{r}(\sin{\theta}-\cos{\theta}){\partial}_{s}\textbf{B}_{1}]=-\frac{1}{\lambda}(<\textbf{B}^{0}>.{\nabla})\textbf{v}\label{15}
\end{equation}
Therefore, splitting of this equation along the Frenet frame one
obtains
\begin{equation}
\frac{{B^{s}}_{1}}{<B^{0}>}=-\frac{1}{\lambda}(1+\frac{{{\tau}_{0}}^{2}}{r^{2}})sin{\theta}v_{\theta}\label{16}
\end{equation}
\begin{equation}
\frac{{B^{\theta}}_{1}}{<B^{0}>}=\frac{1}{\lambda}\frac{rtan{\theta}}{{{\tau}_{0}}^{3}}v_{\theta}\label{17}
\end{equation}
and
\begin{equation}
\frac{1}{r}(\sin{\theta}-\cos{\theta})[\frac{{B^{s}}_{1}}{<B^{0}>}{\tau}_{0}+{{\tau}_{0}}^{2}rsin^{2}{\theta}\frac{{B^{\theta}}_{1}}{<B^{0}>}]=
\frac{1}{\lambda}(-v_{\theta}{{{\tau}_{0}}^{2}}{r^{2}}sin{\theta}+v_{s}{\tau}_{0})\label{18}
\end{equation}
By assuming another approximation of weak torsion and performing
algebraic manipulations with those three last equations yields
\begin{equation}
{\tau}_{0}=\frac{4}{\sqrt{2}}\frac{v_{s}}{v_{\theta}}\label{19}
\end{equation}
Since the toroidal and poloidal components of the flow are given by
$v_{s}={\Omega}_{s}R$ and $v_{\theta}={\Omega}_{\theta}r$, where r
and R are respectively the internal cross-section radius and
external torus radius, which are given in the Perm torus dynamo
experiment \cite{1} by $r=0.02m$ and $R=0.08m$, where the torus
Frenet curvature is given by ${\kappa}_{0}={\tau}_{0}=\frac{1}{R}$,
yields a torsion value by
\begin{equation}
{\tau}_{0}=\approx{7.5{\times}{\Omega}^{\theta}Hz^{-1}m^{-1}}\label{20}
\end{equation}
The magnetic fields are given by
\begin{equation}
{{B^{s}}_{1}}(Na)=-R_{m}\frac{v_{\theta}}{v}{<B^{0}>}\label{21}
\end{equation}
which by making use of the Liquid sodium (Na) torus dynamo data for
the applied field of $<B^{0}>\approx{45G}$ and the induced field of
$B_{1}\approx{0,3G}$ and a $R_{m}\approx{16}$, one obtains
\begin{equation}
\frac{v_{\theta}}{v}\approx{1.0{\times}10^{2}}\label{22}
\end{equation}
Note that poloidal velocities induced by torsion of the twisted
small-scale dynamo flow is much higher than the typical velocities
scales. This is interesting cause the torsion seems to be able to
damp toroidal velocities and work as a brake in the torus dynamo
experiment to substitute the mechanical brake of the Perm torus
dynamo experiment. In the large-scale astrophysical dynamos where
typical magnetic Reynolds numbers are $R_{m}\approx{10^{3}}$ this
same computation yields
\begin{equation}
\frac{v_{\theta}}{v}\approx{10^{-3}}\label{23}
\end{equation}
showing clearly that the poloidal flows are much slower than the
typical velocities in the large-scale dynamo flow. This result is
very well known in solar physics \cite{15}.
\section{Random dynamo twisted flows in torus devices}
In this section a straightforward but long analytical computation
shall be performed to solve the remaining random applied field
induced field equation
\begin{equation}
[{\gamma}-{\lambda}{\Delta}](<{B}^{0}>\textbf{t})={\nabla}{\times}(\textbf{v}{\times}<\textbf{B}_{1}>)\label{24}
\end{equation}
This equation yields the following expressions
\begin{equation}
<{B}^{0}>[{\gamma}+{\lambda}(1-\frac{{{\tau}_{0}}^{2}}{r^{2}})+T_{0}
{\tau}_{0})\textbf{t}+
({\tau}_{0}v_{s}-\frac{v_{\theta}}{r}-{{\tau}_{0}}^{2})\textbf{n}+\textbf{b}(1-\frac{{{\tau}_{0}}^{2}}{r^{2}}){{\tau}_{0}}^{2})]=[\frac{1}{r}
\textbf{e}_{r}\cos{\theta}{\beta}^{{\theta}s}+
\textbf{e}_{\theta}{\partial}_{s}{{\beta}^{{\theta}{s}}}]\label{25}
\end{equation}
where
${\beta}^{{\theta}{s}}:=(v^{s}{B^{\theta}}_{1}-v^{\theta}{B^{s}}_{1})$
and $T^{0}:=\frac{1}{r}(\sin{\theta}-\cos{\theta}){\tau}_{0}$. In
the above computations use has been made of the solenoidal
properties of vectors $\textbf{B}$ and $\textbf{v}$ as
\begin{equation}
{\nabla}.\textbf{v}={\nabla}.\textbf{B}=0\label{26}
\end{equation}
which in Riemannian curvilinear coordinates yields
\begin{equation}
{\partial}_{s}{v}_{\theta}=v_{\theta}{{\tau}_{0}}^{2}rsin{\theta}\label{27}
\end{equation}
the same is valid for the magnetic field toroidal component. The
equations above leads to the following expression for the growth
rate
\begin{equation}
{\gamma}=({\tau}_{0}r\sin{\theta}+\cos^{2}{\theta})[v^{s}\frac{{\textbf{B}_{1}}^{\theta}}{<B^{0}>}-v^{\theta}\frac{{\textbf{B}_{1}}^{s}}{<B^{0}>}]\label{28}
\end{equation}
This equation yields the maximum growth rate of the magnetic random
flow as
\begin{equation}
{\gamma}_{max}=({R_{m}}^{-1}+{R_{m}}r){\Omega}^{\theta}\label{29}
\end{equation}
where $sin{\theta}=cos{\theta}=1$ has been used as the maximum of
trigonometric functions. When the Reynolds number is small as in
small-scale dynamos, $R_{m}\approx{16}$, and by considering dynamo
flows close to the dynamo torus torsioned axis (r=0), this
expression reduces to
\begin{equation}
{\gamma}_{max}={R_{m}}^{-1}{\Omega}^{\theta}\label{30}
\end{equation}
which yields ${\gamma}_{max}=6{\times}10^{-2}{\Omega}^{\theta}$.
Just to give an idea of how big this value is, the galactic dynamo
\cite{16}, yields a growth rate of
${\gamma}_{G}\approx{10^{-5}}{s}^{-1}$.
\section{Conclusions}
A simple proposal to substitute divertors and breakings of Perm
torus dynamo experiment, by a torsioned (twisted) flux tube in the
Riemannian context. Though this solution is similar to recent
proposal done by Shukurov et al \cite{10}, the basic advantage of
the present proposal is that it allows analytical solutions, of the
self-induction equation constraint to the weak torsion approximation
of the torus dynamo device, instead of the numerical simulations
undertaken by Shukurov et al. A Riemannian twisted geometry allows
us to further investigate the Lyapunov exponents in a further
generalization of the present model for a thick cross-section torus
instead of the thin tube used here. The future analysis of the
dynamo operator in the case of the thick dynamo torus device allows
us also to preview physical processes that may happen in Perm torus
dynamo experiment. Another advantage of the use of the twisted
dynamo torus instead of the twisted Moebius strip is technological,
since the technology of building a twisted torus device is already
known from the stellarator plasma devices. Another motivation for
the proposed dynamo magnetic flux tube torus device is that the idea
of the magnetic flux tube as a dynamo has already been applied by
discussed and developed by Schuessler \cite{15} in the context of
solar dynamos. Another motivation stems from the work of Wang et al
\cite{16} in the laminar plasma dynamos in cylinders.
\newpage\section{Acknowledgements} I also am deeply indebt to R Ricca for helpful discussions on
the subject of this paper. Financial supports from Universidade do
Estado do Rio de Janeiro (UERJ) and CNPq (Brazilian Ministry of
Science and Technology) are highly appreciated.

\end{document}